\documentclass[12pt,fleqn]{article}
\usepackage{bm}

\begin{document}
\begin{center}

{\bf \Large Physics of Leap Second }

\vspace{1cm}

Takehisa Fujita\footnote{e-mail:
fffujita@phys.cst.nihon-u.ac.jp} and 
Naohiro Kanda\footnote{e-mail:
nkanda@phys.cst.nihon-u.ac.jp}

Department of Physics, Faculty of Science and Technology, 

Nihon University, Tokyo, Japan

\vspace{2cm}

{\Large Abstract}

\end{center}
The physical origin of the leap second is discussed in terms of the new gravity 
model. The calculated time shift of the earth rotation around the sun 
for one year amounts to $\displaystyle{  \Delta T  \simeq 0.621 \ \ s/ year }$. 
According to the data, the leap second correction for one year corresponds to 
$\Delta T \simeq 0.63 \pm 0.03 \ \ s/ year $, which is in perfect agreement 
with the prediction. This shows that the leap second is not originated from 
the rotation of the earth in its own axis. Instead, it is the same physics as 
the Mercury perihelion shift. We propose a novel dating method (Leap Second Dating) 
which enables to determine the construction date of some archaeological objects 
such as Stonehenge.

\vspace{1cm}
\noindent

\newpage

\section{Introduction}
The observation of the Mercury perihelion shift is well known, and it is the advance 
shift which can be described in terms of the time shift as \cite{fk}
$$ \left({\Delta T\over T}\right) \simeq (2-\varepsilon) 
{G^2M^2\over c^2R^4 \omega^2}   \eqno{(1.1)}  $$
where $G$, $M$ $R$, $\omega$ and $\varepsilon$ denote the gravitational constant, 
the mass of the sun, the radius of the Mercury orbit, the angular velocity  and 
the eccentricity, respectively. In this case, 
one may ask a question as to what should be the time shift for the earth. 
We can easily calculate the time shift for the earth rotation around the sun 
for one year
$$ (\Delta T)_{th} \simeq  0.62  \ \ s/year    \eqno{(1.2)}  $$
where $T$ is taken to be $T=365.24 \ day= 3.16 \times 10^{7}\ s $. 
This means that, as long as one agrees with the observed value of 
the Mercury perihelion advance shift, then one should accept the time shift 
of the earth rotation around the sun as given in eq.(1.2). 

In fact, there is a good observation of the earth rotation around the sun, and 
this is expressed in terms of the leap second. Since people have made the 23 times 
corrections of the leap second for 36.5 years, the time shift per year becomes
$$ (\Delta T)_{exp}  \simeq  0.63 \pm 0.03 \ \ s/year  .  \eqno{(1.3)}  $$
This agrees perfectly with the number which is predicted by the new gravity model 
calculation.

\section{New Gravity Model }
Now, we briefly describe how we can obtain eq.(1.1). Here, 
we start from the Dirac equation for fermions in the gravitational 
potential, and it is written as \cite{tfgrav,fujita}
$$ \left[-i \bm{\nabla}\cdot \bm{\alpha}+ \left(m -{GmM\over r}\right) 
\beta \right] \psi =E\psi . \eqno{(2.1)} $$
It is essential that we should write the Dirac equation under the gravitational 
force. Otherwise, we cannot obtain the Newton equation starting from the most 
fundamental equation of motion. The first step is to reduce the Dirac equation 
to the non-relativistic form of the equation. This is of course straightforward, and 
one can obtain the non-relativistic equation of motion 
by making use of the Foldy-Wouthuysen transformation \cite{bd}. Further, as the second 
step, we can make the classical limit of the Schr\"odinger equation, which is also 
a straightforward calculation. In this case, we obtain a new gravitational potential 
$$ V(r)= - {GmM\over r} +{1\over 2mc^2}\left({GmM\over r}\right)^2 . \eqno{(2.2)}  $$
The Newton equation with the new gravitational potential can be written as
$$ m \ddot{r} = -{GmM\over r^2} +{\ell^2\over mr^3} +{G^2M^2m\over c^2r^3}  
 \eqno{(2.3)}  $$
where $\ell =m r^2 \dot{\varphi} $. 
The orbit solution of eq.(2.3) can be written as 
$$ r={A\over{1+ \varepsilon \cos \left( \varphi(1+{1\over 2}\eta) 
\right) }}  \eqno{(2.4)}  $$
where $A$ and $\varepsilon$ are given as
$$ A={\ell^2\over GMm^2 }(1+\eta), \ \ \ \ 
\varepsilon=\sqrt{1+{2\ell^2(1+\eta)E\over m(GmM)^2}}.  \eqno{(2.5)}  $$
Here,  $\eta$ is defined as
$$ \eta =  {G^2M^2\over c^2R^4 \omega^2}   \eqno{(2.6)}  $$
where the angular velocity $\omega$ and radius $R$ are defined as
$$ \omega \equiv {\ell\over mR^2}, \ \ \ \ 
R \equiv {\ell^2\over GMm^2(1-\varepsilon^2)^{3\over 4}} .  \eqno{(2.7)}  $$
Physical observables can be obtained by integrating 
$\dot{\varphi}= {\ell\over mr^2}$ over the period $T$
$$ {\ell\over m}\int_0^T dt = \int_0^{2\pi} r^2 d\varphi 
= A^2  \int_0^{2\pi} {1\over{\left(1+ \varepsilon \cos 
\left( {L\over \ell}\varphi \right)  \right)^2}} d\varphi.  \eqno{(2.8)}  $$
This can be easily calculated to be 
$$ \omega T=2\pi(1+2\eta)\left( 1-\varepsilon\eta \right)
\simeq 2\pi\{ 1+(2-\varepsilon) \eta\}  \eqno{(2.9)}  $$
where $\varepsilon$ is assumed to be small. 
Therefore, the new gravity potential gives rise to the advance shift 
of the period $T$, and it can be written as 
$$ \left({\Delta T\over T}\right)_{th} \simeq (2-\varepsilon)\eta . 
 \eqno{(2.10)}  $$
This is the time shift which should be compared to experiment. 
It should be important to note that the calculated effect must be there, regardless 
the presence of other effects. Indeed, for the earth rotation around the sun, 
there may be some other effects such as the effects of the moon or other planets. 
However, as long as we carry out naive estimations of the moon effect, we see that 
the effects are wiped away because of the averaging over the motion.  But still it 
may be interesting to calculate these effects in future as the corrections to the new 
gravitational potential effect. 

\section{Time Shift of Earth Rotation $-$ Leap Second }
Here, we calculate the time shift of the earth rotation around the sun. 
First, we evaluate the $\eta$ 
$$ \eta = {G^2M^2\over c^2R^4 \omega^2} \simeq 0.992 \times 10^{-8}  \eqno{(3.1)}  $$
where we employ the following values for $R$, $M$ and $\omega$
$$ R=1.496 \times 10^{11} \ {\rm m}, \  M= 1.989 \times 10^{30} \ {\rm kg}, 
\  \omega=1.991 \times 10^{-7} .   $$
In this case, we find  the time shift for one year
$$ (\Delta T)_{th}  \simeq  0.621  \ \ s/year   \eqno{(3.2)}  $$
where $\varepsilon =0.0167 $ is taken. 
In fact, people have been making corrections for the leap second, 
and according to the data, they made the first leap second correction in June of 1972. 
After that, they have made the leap second corrections from December 1972 
to December 2008. The total corrections amount to 23 seconds for 36.5 years 
since we should start from June 1972. This corresponds to the time shift per year
$$ (\Delta T)_{exp}  \simeq 0.63 \pm0.03 \ \  s/year   \eqno{(3.3)}  $$
where the errors are supposed to come from one year shift of the observation. 
This agrees surprisingly well with the theoretical time shift of the earth.

\section{Prediction from General Relativity }
Here, we discuss the calculated result by the general relativity \cite{ein,misner}. 
People only considered the shift due to the angular variable $\varphi$ as 
$$ \cos \varphi \longrightarrow \cos (1-\gamma) \varphi   \eqno{(4.1)} $$
where $\gamma$ is found to be 
$$ \gamma = {3G^2M^2\over c^2R^4 \omega^2} . \eqno{(4.2)}  $$
In this case, one finds the physical observable
$$ \omega T \simeq 2\pi( 1+2\varepsilon \gamma ) . \eqno{(4.3)}  $$
Therefore, the time shift predicted by the general relativity becomes
$$ (\Delta T)_{th}  \simeq  0.031  \ \ s/year  . \eqno{(4.4)}  $$
This shows that it is much smaller 
than the observed time shift of the earth. 

In reality, if the angular momentum is affected from the external 
potential as given in eq.(4.1), then not only the angular variable 
but also $A$ in eq.(2.5) are changed, and therefore as the total effects of 
the physical observables, we find
$$ \omega T \simeq 2\pi\{ 1-2(2-\varepsilon) \gamma \}  \eqno{(4.5)}  $$
which, unfortunately, gives a retreat shift. 
In this respect, the prediction of the general relativity disagrees 
with all the data observed in the Mercury perihelion shift, the GPS time shift 
and the leap second correction of the earth \cite{fk}.

\section{Leap Second  Dating }
Since we know quite accurately the time shift of the earth rotation around 
the sun by now, we may apply this time shift to the dating of some archaeological 
objects such as pyramids or Stonehenge. For example, the time shift of 1000 years 
amounts to 10.3 minutes, and some of the archaeological objects may well possess 
a special part of the building which can be pointed to the sun at the equinox. 
In this case, one may be able to find out the date when this object was constructed. 
This new dating procedure is basically useful for the stone-made archaeological objects
in contrast to the dating of the wooden buildings which can be determined 
from the Carbon dating. It should be noted that the new dating method has an important 
assumption that there should be no major earthquake in the region of the archaeological 
objects. 


\end{document}